\documentclass{article}

\usepackage{amsmath,amssymb,amsfonts}
\usepackage{amsthm,latexsym}
\usepackage{bbm}
\usepackage{graphicx}
\usepackage{cite}

\newtheorem{theor}{Theorem}[section]
\newtheorem{lem}{Lemma}[section]
\newtheorem{prop}{Proposition}[section]
\newtheorem{corol}{Corollary}[section]
\newtheorem{rem}{Remark}[section]

\renewcommand{\theequation}{\thesection.\arabic{equation}}
\numberwithin{equation}{section}

\def\eps{\varepsilon}
\def\R{{\mathbb R}}
\def\C{{\mathbb C}}

\def\sh{\mathrm{sh}\,}
\let\int=\intop
\let\phi=\varphi

\def\Ln{\mathrm{Ln}\,}
\def\nint{\diagup \!\!\!\!\! \!\!\int}

\DeclareMathOperator*{\res}{res}

\title{On point-like interaction of three particles:
two fermions and another particle. II}

\author{R.A. Minlos\thanks{Institute for Information Transmission
Problems of Russian Academy of Sciences,
Bolshoy Karetnyi 19, Moscow, Russia.
E-mail: minl{@}iitp.ru} }

\date{November 1, 2012}

\begin{document}

\maketitle

\begin{abstract}
This work continues \cite{bib1} where the construction of
Hamiltonian $H$ for the system of three quantum particles is considered.
Namely the system consists of two fermions with mass $1$ and another
particle with mass $m>0$.
In the present paper, like in \cite{bib1}, we study the part $T_{l=1}$
of auxilliary operator $T = \oplus_{l=0}^{\infty} T_l$
involving the construction of the resolvent for the operator $H$.
In this work together with the previous one two constants
$0<m_1<m_0<\infty$ were found such that:
1) for $m>m_0$ the operator $T_{l=1}$ is selfadjoint
but for $m \leqslant m_0$ it has the deficiency indexes $(1,1)$;
2) for $m_1<m<m_0$ any selfadjoint extension of $T_{l=1}$ is semibounded
below; 
3) for $0<m<m_1$ any selfadjoint extension of $T_{l=1}$ has the sequence 
of eigenvalues $\{ \lambda_n <0, n> n_0\}$ with the asymptotics
\[
\lambda_n = \lambda_0 e^{\delta n } + O(1),\quad n\to\infty,
\]
where $\lambda_0 <0$, $\delta >0$, $n_0>0$ and there is'nt other spectrum  
on the interval $\lambda < \lambda_{n_0}$.
\end{abstract}


\section{Introduction and main results}

This paper is continuation of work \cite{bib1}
(see also \cite{bib2,bib3,bib4,bib5})
devoted to construction of Hamiltonian
for the system of three point-like interacting
particles: two fermions with mass $1$
and a different particle with mass $m>0$.
In the mentioned papers the construction of Hamiltonian begins with the 
introduction of the symmetric operator
\begin{equation}
\label{eq1.1}
H_0 = -\frac 12 \Big( \frac 1m \Delta_y +
\Delta_{x_1} + \Delta_{x_2} \Big)
\end{equation}
where $x_1$, $x_2 \in \R^3$ are the positions of the fermions,
$y \in \R^3$ is the position of another particle, $\Delta_y$,
$\Delta_{x_1}$, $\Delta_{x_2}$ are Laplacians w.r.t.\ these variables.
The operator $H_0$ is given on the set $D(H_0)\subset L_2 (\R^3)
\otimes L_2^{\mathit{asym}}(\R^3 \times \R^3)$ consisting of smooth functions
$\Psi(y,x_1,x_2)$ rapidly decreasing at infinity, antisymmetric
w.r.t.\ variables $x_1$, $x_2$ and satisfying the following conditions:
\begin{equation}
\label{eq1.2}
\Psi(y,x_1,x_2) \big|_{y=x_i} =0, \quad i=1,2.
\end{equation}
The operator $H_0$ has non-zero deficiency indexes and ``true''
Hamiltonian is contained among its selfadjoint extensions.
Usually a one-parametric family $\{H_\eps,\eps\in
\R^1\}$ of such extensions is considered, the so-called
Ter-Martirosian--Scornyakov's extensions (see \cite{bib6,bib7}).
However,  for some values of parameter $m$ this extension is not
selfadjoint.

In such cases a problem of description of selfadjoint extensions
for $H_\eps$ arises.  It is shown in 
\cite{bib4,bib5} that this problem is equivalent to the one for some auxiliary symmetric 
operator $T$ acting in the space $L_2(\R^3)$ according to the formula:
\begin{align}
\label{eq1.3}
(Tf)(k) =&\,
2 \pi^2 \bigg( \sqrt{ \Big( 1 - \Big(\frac \mu2 \Big)^2 \Big)
k^2 + 1} \,  f(k) \\
& {}+
\int_{\R^3} \frac{f(p)\, dp}{k^2+p^2+\mu(p,k)+1} \bigg) ,
\quad k\in\R^3 \nonumber
\end{align}
$(\mu = 2/(m+1)$). This operator is defined on the set
\begin{equation}
\label{eq1.4}
D(T) = \big\{ f\in L_2(\R^3): |k|f(k) \in L_2(\R^3)\big\}.
\end{equation}
In addition, every selfadjoint extension of $T$ generates by specific
way a selfadjoint extension of $H_\eps$ (see \cite{bib4}) and so all
extensions of $H_\eps$ are formed. In particular any eigenvalue
$\lambda$ of selfadjoint extension $\hat T$ of the operator $T$
generates a negative eigenvalue of the corresponding extension of
$H_\eps$ that is equal to
\begin{equation}
\label{eq1.5}
-\Big( \frac \lambda\eps \Big)^{-2} .
\end{equation}
The operator $T$ commutes with the operators of representation of
rotation group~$O_3$:
\begin{equation}
\label{eq1.6a}
(\tau_g f)(k) = f(g^{-1}k), \quad g\in O_3
\end{equation}
acting in the space $L_2(\R^3)$. Thus the operator $T$ is decomposed
in the direct sum of the operators $T_l$, $l=0,1,{\ldots} $ acting
each in the subspace
\begin{equation}
\label{eq1.6b}
\mathcal H _l = L_2(\R^1_+, r^2\, dr) \otimes L_2^l(S) \subset L_2(\R^3)
\end{equation}
correspondingly.
Here $ L_2^l(S) \subset L_2(S)$ is the subspace of functions on a unit
sphere $S\in \R^3$ where the irreducible representation of the group
$O_3$ with weight $l$ acts (subspace of the spherical functions of
weight $l$, see \cite{bib7}).
The operator $T_l$ is tensor product of operators:
\begin{equation}
\label{eq1.7}
T_l = M_l \otimes E_l,
\end{equation}
where $E_l$ is unit operator in $L_2^l(S)$ and the symmetric operator
$M_l$ acts in the space $L_2(\R_1^+, r^2\,dr)$ by formula
\begin{align}
\label{eq1.8}
(M_l \phi) (r) = 2 \pi^2 \bigg( \sqrt{ \Big( 1 - \Big( \frac \mu2
\Big)^2 \Big) r^2 +1} + \int_{-1}^1 P_l(x)\, dx \int_0^\infty
\frac{(r')^2 \phi(r') \, dr'}{(r')^2 + r^2 + \mu x r r'} \bigg)
\end{align}
where $\{ P_l(x), l=0,1,{\ldots} \}$ are polynoms of Legendre with
conditions: $P_l(1)=1$. The domain of $M_l$ is
\begin{align}
\label{eq1.9}
D(M_l) = \{\phi \in L_2(\R_1^+, r^2\,dr) :
r \phi(r) \in L_2(\R_1^+, r^2\,dr) \} .
\end{align}
Note that every selfadjoint extension of $T$ commuting with the
operators $\tau_g$ (see \eqref{eq1.6a}) generates a selfadjoint
extension of each operator $M_l$ and vice versa any collection of
selfadjoint extensions of all operators $M_l$ reduces to the one for the
operator $T$ (commuting with $\tau_g$, $g\in O_3$).

In \cite{bib1} the operators $M_{l=0}$ and $M_{l=1}$ were studied
in detail and it was proved that the operator $M_{l=0}$ is selfadjoint
and semibounded below for all values of $m$ while for the operator
$M_{l=1}$ there exists a constant $m_0$ such that for $m>m_0$ this operator
is selfadjoint and bounded below but for $m<m_0$ the operator
$M_{l=1}$ has deficiency indexes $(1,1)$. In this paper these results
are complemented by the next assertion:
\begin{theor}
\label{thm1.1}
There exists a constant $0<m_1<m_0$ so that:
\begin{itemize}
\item[{\rm 1.}]
for $m\in[m_1,m_0)$ the operator $M_{l=1}$ is bounded  below,
\begin{align}
\label{eq1.10}
\inf_{\|\phi\|=1} (M_{l=1} \rho, \phi) = k > -\infty ,
\end{align}
and the part of spectrum of any selfadjoint extension $M_{l=1}^\beta$
of the operator $M_{l=1}$ (the parameter $\beta$, $|\beta|=1$, is
introduced below) lying below $k$ consists of not more than one
eigenvalue;
\item[{\rm 2.}]
for $m\in(0,m_1)$ the operator $M_{l=1}$ is not bounded below and any
its selfadjoint extension $M_{l=1}^\beta$ has the sequence of
eigenvalues $\{ \lambda_n, n>n_0\}$ going off to $-\infty$ with the
asymptotics
\begin{align}
\label{eq1.11}
\lambda_n = \lambda_0 e^{n \delta} + O(1), \quad n\to\infty
\end{align}
where $\lambda_0 <0$ and $n_0$ depends on the  extension, but $\delta >0$
is the same for all extensions;
\item[{\rm 3.}]
for every selfadjoint extension $M_{l=1}^\beta$ there exists a 
constant $\kappa=\kappa(\beta)<0$ so that the part of spectrum of
$M_{l=1}^\beta$ lying below $\kappa$ consists only of the elements of
the sequence \eqref{eq1.11}.
\end{itemize}
\end{theor}

\begin{rem}
Evidently, an eigenvalue $\lambda$ of any selfadjoint extension of
$M_{l=1}$ is a triple eigenvalue of the corresponding extension of the
operator $T_{l=1}$ and hence of some extensions of the operator $T$. From
here it follows that there exist selfadjoint extensions of
$H_\eps$ having a sequence of eigenvalues of form
\begin{align}
\label{eq1.12}
-\Big( \frac{\lambda_n}\eps \Big)^{-2}, \quad n>n_0,
\end{align}
as it follows from \eqref{eq1.5} ($\lambda_n$ are defined by 
 \eqref{eq1.11}). The sequence \eqref{eq1.12}  approaches 
zero for $n\to\infty$ i.e.\ the Efimov's effect appears.
\end{rem}

\section{Some reminders and beginning of the proof of Theorem \ref{thm1.1}}

Here we explain some facts from the previous works (see
\cite{bib1,bib5}).

\medskip
\noindent{\bf 1.}
The operator $M_{l=1}$ can be represented in form 
\begin{align}
\label{eq2.1}
M_{l=1} = M_0+M_1
\end{align}
where $M_1$ is a bounded selfadjoint operator and $M_0$ is the
symmetric operator acting by the formula
\begin{align}
\label{eq2.2}
(M_0 \phi) (r) = 2 \pi^2  \bigg( \sqrt{ 1 - \Big( \frac \mu2 \Big)^2 }
r \phi(r) + \int_{-1}^1 P_1(x)\, dx \int_0^\infty
\frac{(r')^2 \phi(r') \, dr'}{(r')^2 + r^2 + \mu x r r'} \bigg)
\end{align}
on the domain $D(M_0) = D(M_{l=1})$.

Below we shall study mainly the operator $M_0$ returning from time to time
 to the operator $M_{l=1}$.

\medskip
\noindent{\bf 2.}
Denote by $\mathcal J$ the set of functions $F(z)$ defined in the closed
strip
\begin{align}
\label{eq2.3}
I = \{ z \in \C^1 : 0 \leqslant \Im z \leqslant 1 \},
\end{align}
continuous and analytical inside $I$, satisfying the condition
\begin{align}
\label{eq2.4}
\sup_{0\leqslant t \leqslant 1} \int_{-\infty}^{\infty}
| F(s+it)|^2\, ds < \infty .
\end{align}
The contraction of the function $F\in \mathcal J$ on the lines
\begin{align*}
l_+ = \{ z=s,s\in \R^1\}, \quad
l_- = \{ z=s+i,s\in \R^1\}
\end{align*}
limiting the strip $I$ is denoted by $F_+(s)$ and $F_-(s)$
correspondingly. Denote by $j\in L^2(\R^1)$ the set of functions $f(s)
\in L^2(\R^1)$ which have the form
\begin{align}
\label{eq2.5}
f(s) = F_+(s), \quad s\in \R^1
\end{align}
for some (and unique) function $F\in \mathcal J$. This function
$F=F(f)$ is called {\it superstucture\/} on $f$.

Consider unitary Mellin's map
\begin{align}
\label{eq2.6a}
\omega : L_2(\R_1^+, r^2\,dr) \longrightarrow L_2(\R^1, ds) :
\phi \longrightarrow f=f(s) = \frac{1}{\sqrt{2\pi}} \int_0^{\infty}
r^{-is+1/2} \phi(r) \, dr
\end{align}
and the inverse map:
\begin{align}
\label{eq2.6b}
\omega^{-1} : f \longrightarrow \phi = \phi(r) = \frac{1}{\sqrt{2\pi}}
\int_{-\infty}^{\infty} r^{is-3/2} f(s)\, ds .
\end{align}
The map $\omega$ transfers the domain $D(M_0)$ of the operator $M_0$
in the set $j : \omega (D(M_0))=j$. In addition the operator $M_0$ is
passed by $\omega$ to a unitary equivalent operator $\tilde{M}_0$:
\begin{align}
\label{eq2.7}
\tilde{M}_0 = \omega M_0 \omega^{-1}
\end{align}
acting in the space $L_2(\R^1,ds)$
with the help of the formula
\begin{align}
\label{eq2.8}
\big( \tilde{M}_0 f \big) (s) = (N(z) F(z))_- (s) = N_- (s) F_- (s),
\quad f \in D(\tilde{M}_0)
\end{align}
where $F=F(f)$ is superstructure on $f$ and $N(z)$ is analytical
function defined in the strip $I$:
\begin{align}
\label{eq2.9}
N(z) = 2\pi^2 \Big( \sqrt{ 1 - \Big(\frac \mu2 \Big)^2 }
- \Lambda(z) \Big)
\end{align}
where
\begin{align}
\label{eq2.10}
\Lambda(z) = \int_0^1 x \, dx \, \frac{ \sh( v(x)(z-i/2)) }{
\cos (v(x)) \cdot \sh ((\pi/2)(z-i/2)) }
\end{align}
and $v(x) = \arcsin \mu x/2$.

Note that the function $N(z)$ is invariant w.r.t.\ the reflection
 of the strip $N(z)$ into itself:
\begin{align}
\label{eq2.11}
z \rightarrow z^* = -z+i.
\end{align}

\medskip
\noindent{\bf 3.}
It is shown in \cite{bib1} that the operator $\tilde{M}_0$ is
selfadjoint for those values of $\mu = 2/(m+1)$ which lie 
in the interval $(0,\mu_0)$ where $\mu_0$ is a unique root of the equation
\begin{align}
\label{eq2.12}
\sqrt{ 1 - \Big(\frac \mu2 \Big)^2 } = \Lambda(0) \equiv q_0(\mu).
\end{align}
For $\mu \geqslant \mu_0$ the function $N(z)$ has two simple zeros
$z_\pm \in I$ and the operator $\tilde{M}_0$ is not selfadjoint; it
has deficiency indexes $(1,1)$. Further, for $\mu \in (\mu_0,\mu_1)$,
where $\mu_1 > \mu_0$ is a unique root of the equation
\begin{align}
\label{eq2.13}
\sqrt{ 1 - \Big(\frac \mu2 \Big)^2 } = \Lambda \Big( \frac i2 \Big)
\equiv q_1(\mu),
\end{align}
the zeros $z_\pm$ lie on the unit interval of imaginary axis
$\tau=(0,i)$:
\begin{align}
\label{eq2.14}
z_\pm = \frac i2 \pm i t_0, \quad t_0 = t_0(\mu) \in \Big( 0, \frac 12
\Big].
\end{align}
For $t_0 = 0$ these zeros coincide, forming a unique zero of 
$N(z)$ with multiplicity $2$. For $\mu \in (\mu_1, 2)$ the zeros
$z_\pm$ lie on the line
\begin{align*}
l_{1/2} = \{ z = s +i/2, s \in \R^1\}
\end{align*}
and have the form
\begin{align}
\label{eq2.15}
z_\pm = \frac i2 \pm s_0, \quad s_0 = s_0(\mu) \in (0,\infty).
\end{align}
The graphs of functions $\sqrt{ 1 - (\mu/2)^2}$, $q_0(\mu)$ and
$q_1(\mu)$ are presented in Figure~\ref{fig1}.

\begin{figure}[ht]
\centering{
\begin{picture}(180,85)
\put(0,10){\line(1,0){170}}
\qbezier(20,70)(40,11)(150,11)
\qbezier(150,50)(90,11)(20,11)
\qbezier(150,90)(90,11)(20,11)
\put(20,10){\line(0,1){60}}
\put(150,10){\line(0,1){80}}
\put(17,0){$0$}
\multiput(81,10)(0,4){2}{\line(0,1){2}}
\put(81,0){$\mu_1$}
\multiput(71,10)(0,4){3}{\line(0,1){2}}
\put(64,0){$\mu_0$}
\put(148,0){$2$}
\put(25,65){\rotatebox{-30}{$\sqrt{1-(\mu/2)^2}$}}
\put(125,47){\rotatebox{15}{$q_1(\mu)$}}
\put(115,67){\rotatebox{30}{$q_0(\mu)$}}
\end{picture}
}
\caption{\label{fig1}}
\end{figure}

\medskip
\noindent{\bf 4.}
Note that
\begin{align}
\label{eq2.16}
\min_{z\in \tau} \Lambda (z) = \Lambda (i/2) = \max_{z\in l_{1/2}}
\Lambda (z)
\end{align}
and consequently, for $\mu \in (\mu_0,\mu_1]$, for  $\Lambda (i/2)
\leqslant \sqrt{1- (\mu/2)^2}$,
\begin{align}
\label{eq2.17}
N(z) \big|_{l_{1/2}} = N \Big( \frac i2 +s \Big) \geqslant 0 .
\end{align}
On the other hand, as it follows from the construction in \cite {bib1}
the quadratic form $(\tilde{M}_0 f, f)$ can be represented as
\begin{align}
\label{eq2.18}
(\tilde{M}_0 f, f) = \int_{-\infty}^{\infty} \Big| F \Big( \frac i2 +s
\Big) \Big|^2 N\Big( \frac i2 +s \Big) \, ds , \quad
f \in D(\tilde{M}_0)
\end{align}
where $F=F(f)$ is superstructure on $f$. From \eqref{eq2.17} and
\eqref{eq2.18} it follows that the operator $\tilde{M}_0$ (and also
${M}_0$) is non-negative:
\begin{align}
\label{eq2.19a}
(\tilde{M}_0 f, f) \geqslant 0, \quad \big( (M_0 \phi, \phi)
\geqslant 0 \big), \quad f \in D(\tilde{M}_0), \quad
\phi \in D(M_0) .
\end{align}
Returning to the symmetric operator $M_{l=1}$ we find from 
\eqref{eq2.1} and~\eqref{eq2.19a} that
\begin{align}
\label{eq2.19b}
\inf_{\|\phi\|=1} (M_{l=1}\phi, \phi) \geqslant \inf_{\|\phi\|=1}
(M_1\phi, \phi) \equiv k > -\infty.
\end{align}
It is known that if a symmetric operator with the deficiency
indexes $(1,1)$ is bounded below by a constant $k$, the part of spectrum
for any its selfadjoint extension that lies below $k$ consists of not more
than one eigenvalue. Setting $m_1 = 2/\mu_1 -1$ and $m_0 = 2/\mu_0 - 1$
we arrive to the first assertion of Theorem \ref{thm1.1}.

\medskip\noindent{\bf 5.}
For $\mu > \mu_0$, $\mu \neq \mu_1$, the domain $D(\tilde{M}_0^*)$ of the
operator $\tilde{M}_0^*$ consists of the functions of the form
\begin{align}
\label{eq2.20a}
g(s) = f(s) + \frac{C_+}{s-z_+} + \frac{C_-}{s-z_-}
\end{align}
where $f \in D(\tilde{M}_0)$ and $C_\pm$ are arbitrary
constants. In the case $\mu = \mu_1$, when $z_+ = z_- = z_0$
\begin{align}
\label{eq2.20b}
g(s) = f(s) + \frac{C_1}{s-z_0} + \frac{C_2}{(s-z_0)^2} .
\end{align}
The meromorphic function $G(z)$, $z\in I$ of the form
\begin{align}
\label{eq2.21a}
G(z) = F(z) + \frac{C_+}{z-z_+} + \frac{C_-}{z-z_-}
\end{align}
in the case \eqref{eq2.20a} or
\begin{align}
\label{eq2.21b}
G(z) = F(z) + \frac{C_1}{z-z_0} + \frac{C_2}{(z-z_0)^2}
\end{align}
in the case \eqref{eq2.20b} is called the superstructure on $g$. Here
the function $ F(z)$ is the superstructure on $f$. The operator
$\tilde{M}_0^*$ acts on $D(\tilde{M}_0^*)$ as before by the formula
\begin{align}
\label{eq2.22}
(\tilde{M}_0^* g) (s) = N(z) \, G(z) \big|_{z=s+i} = N_-(s) \, G_-(s)
\end{align}
where $G=G(g)$ is superstructure on $g$. The set of functions of
the form~\eqref{eq2.21a} is denoted by $\mathcal{J}^*$. It is easy to prove
the following assertion:

\begin{prop}\label{prop2.1}
The meromorphic function $G(z)$,
$z\in I$ with the simple poles $z_+$, $z_-$ belongs to the class
$\mathcal{J}^*$ if there exists some bounded open
set $U\in I$ containing $z_+$ and $z_-$ so that
\begin{align}
\label{eq2.23}
\sup_{0 \leqslant t \leqslant 1} \int_{-\infty}^{\infty} \chi_{I
\setminus U}
(s+it) \, | G(s+it) |^2 \, ds < \infty
\end{align}
where $\chi_{I \setminus U}$ is characteristic function of the set
${I \setminus U}$. And
conversely if $G \in \mathcal{J}^*$, the inequality \eqref{eq2.23} is
true for any bounded open set $U\in I$ containing $z_+$ and~$z_-$.
\end{prop}

\section{Selfajoint extensions of the operator $\boldsymbol{M}$ for
$\boldsymbol{m\in(0,m_1)}$ and its negative spectrum}

Above we have studied the operator $\tilde{M}_0$ and also $M_{l=1}$
for the case $m\in(m_1,m_0)$. We now turn to the case $m\in(0,m_1)$. At
first we shall construct the eigenvectors of the operator
$\tilde{M}_0^*$ with the eigenvalues $\lambda$ lying outside  the
positive semiaxis $(0,\infty)$:
\begin{align}
\label{eq3.1}
\lambda = |\lambda| e^{i\theta}, \quad 0 <\theta< 2\pi .
\end{align}
These eigenfunctions $g^\lambda$ satisfy the equation
\begin{align}
\label{eq3.2a}
(\tilde{M}_0^* g^\lambda ) (s) = \lambda g^\lambda (s)
\end{align}
which, proceeding to the superstructure $G^\lambda$ on $g^\lambda$, 
takes the form
\begin{align}
\label{eq3.2b}
N_- (s) G^\lambda_- (s) = \lambda G^\lambda_+ (s) .
\end{align}
For the solution of this equation and further investigation of the
functions $G^\lambda$ we perform a new reduction of our picture,
namely we change the variables:
\begin{align}
\label{eq3.3}
w = e^{2\pi z}, \quad z \in I.
\end{align}
While the variable $z$ runs over the strip $I$, the variable $w$ varies 
in the set $\bar{\Pi}$ consisting of the area $\Pi = \C^1 \setminus
[0,\infty)$ and two coasts of the cut $[0,\infty)$ -- the upper one,
$\hat{l}_+$, and the lower one, $\hat{l}_-$. In addition, the boundary of
the strip $I$ consisting of two lines, ${l}_+$ and ${l}_-$,  proceeds to the
boundary of $\Pi$: $l_\pm \to \hat{l}_\pm$. For any function $H(z)$ 
defined in the strip $I$, we denote its values w.r.t.\ the variable $w$ by 
 $\hat{H} (w)$:
\begin{align}
\label{eq3.4}
\hat{H} (w) = H \Big( \frac{\ln w}{2\pi} \Big).
\end{align}
Here $\ln w$ means the branch of the logarithmic function continuous in
$\bar{\Pi}$ with imaginary part included in $(0,2\pi]$. The boundary
values of $\hat{H} (w)$ on the coasts $l_\pm$ are denoted by
$H_\pm (t)$ or $H_\pm (t\pm i0)$. The Hilbert space $L_2(\R^1,ds)$ of the
functions on the line $l_+$ passes to the space $L_2(\R_+^1,dt/t)$ of the
functions on the coast $\hat{l}_+$. The functions $F(z) \in
\mathcal{J}$ pass to the functions $\hat{F} (w)$ continuous in
$\bar{\Pi}$, analytical in $\Pi$ and satisfying a condition similar
to \eqref{eq2.4}:
\begin{align}
\label{eq3.5}
\sup_{0 \leqslant \psi \leqslant 2\pi} \int_0^\infty \frac{
| \hat{F} (r e^{i\psi}) |^2 }{r} \, dr < \infty .
\end{align}
The set of such functions is denoted by $\hat{\mathcal{J}}$.
Analogously the functions $G\in \mathcal{J}^*$ pass to the meromorphic
functions $\hat G$ on $\bar\Pi$ with the simple poles $w_\pm$ and
satisfying a condition similar to \eqref{eq2.2}
\begin{align}
\label{eq3.6}
\sup_{ 0 \leqslant \psi \leqslant 2\pi} \int_0^\infty \chi_{\bar{\Pi} \setminus U}
(r e^{i\psi} ) \frac{ | \hat{G} (r e^{i\psi}) |^2 }{r} \, dr < \infty .
\end{align}
Here $w_\pm = \exp\{2\pi z_\pm \} = -\exp\{ \pm 2\pi s_0\}$,
$U\in \bar\Pi$ is a bounded area containing the points $w_\pm$ and
$\chi_{\bar{\Pi} \setminus U}$ is characteristic function of $\bar{\Pi}
\setminus U$. The set of such functions $\hat G$ is denoted by
$\hat{\mathcal{J}}^*$. The domains of the operators $\tilde{M}_0$ and
$\tilde{M}_0^*$ pass to the sets of functions $\hat f$ and $\hat g$
of the form
\begin{align}
\label{eq3.7}
\hat f (t) = \hat F_+ (t+i0) \equiv F(w) \big|_{\hat{l}_+}, \quad
\hat F \in \hat{\mathcal{J}}, \\
\hat g (t) = \hat G_+ (t+i0) \equiv G(w) \big|_{\hat{l}_+}, \quad 
\hat G \in \hat{\mathcal{J}}^* . \nonumber
\end{align}
As before, the functions $\hat F = \hat F (\hat f)$ and $\hat G = \hat G (\hat g)$
are called the superstructures on~$\hat f$ and $\hat g$
correspondingly. The operators $\tilde{M}_0$ and $\tilde{M}_0^*$ pass
due to the change \eqref{eq3.3} to the operators $\hat{M}_0$ and
$\hat{M}_0^*$ acting by formula
\begin{align}
\label{eq3.8}
( \hat{M}_0 \hat f ) (t+i0) = \hat{N}_- (t-i0) \hat{F}_- (t-i0), \quad
\hat f \in D(\hat{M}_0)
\end{align}
and the same for $\hat{M}_0^*$.

The equation \eqref{eq3.2b} for $G^\lambda$ now reads for
$\hat{G}^\lambda$ 
\begin{align}
\label{eq3.9}
(\hat{N}_-^* \hat{G}^\lambda_- ) (t-i0) = \lambda^* G^\lambda_+ (t+i0)
\end{align}
where
\begin{align*}
\hat{N}^* (w) = \frac 1{2 \pi^2 \sqrt{ 1 - (\mu/2)^2 } }\cdot \hat{N}
(w) = 1 - \frac 1{\sqrt{ 1 - (\mu/2)^2 }} \, \hat\Lambda (w)
\end{align*}
and
\begin{align*}
\lambda^* = \frac \lambda{2 \pi^2 \sqrt{ 1 - (\mu/2)^2 } } .
\end{align*}
We are going to write the explicit solution of equation \eqref{eq3.9} but
first we need to introduce some quantity.

\medskip
\noindent{\bf 1.}
Denote by $ h (w)$ the following function on $\bar\Pi$
\begin{align}
\label{eq3.10}
h (w) = \frac 1{2 \pi} \ln w - s_0 - \frac i2 .
\end{align}
It is obvious that $ h (w_+) = 0$ and $w_+$ is a unique zero of
$ h (w)$.

\medskip
\noindent{\bf 2.}
Introduce then a function $a(x)$, $x \in (0,\infty)$, by 
\begin{align}
\label{eq3.11}
a(x) = \hat{N}_-^* (x) \frac{{h}_+ (x)}{h_- (x)} .
\end{align}
This function is smooth and distinct from zero everywhere. Indicate
the following properties of this function.

\begin{lem}\label{lem3.1}
The increment of $\arg a(x)$ for the change of $x$ from zero to infinity
is equal to zero:
\begin{align}
\label{eq3.12}
\Delta_{x=0}^{x=\infty} \arg (a(x)) = 0.
\end{align}

The function $\Ln (a(x))$ has the following asymptotics as $x\to\infty$:
\begin{align}
\label{eq3.13}
\Ln (a(x)) = -\frac{2\pi i}{\ln x} + O\Big( \frac 1{ (\ln x)^2} \Big).
\end{align}
Here $\Ln u$, $u \in \C^1 \setminus (-\infty,0)$ is the branch of the
logarithmic function defined on the complex plane with the cut
$(-\infty,0)$ with imaginary part constrained between $-i\pi$ and $i\pi$.
\end{lem}

The proof of this lemma is contained in Appendix.

\begin{corol}
From \eqref{eq3.12} and \eqref{eq3.13} it follows that for any
$w\in\Pi$ there exists the limit
\begin{align}
\label{eq3.14}
\lim_{n\to\infty} \bigg( \frac 1{2\pi i} \int_0^n \frac {\Ln (a(x)) }
{x-w} \, dx + \Ln (\ln n) \bigg) \equiv K(w)
\end{align}
(a regularized integral) which we shall write in the form
\begin{align}
\label{eq3.15}
K(w) = \frac 1{2\pi i} \, \nint_0^\infty \, \frac {\Ln (a(x)) }
{x-w} \, dx, \quad w\in\Pi .
\end{align}
\end{corol}

It is evident that $K(w)$ is analytical function in $\Pi$, then because of
the smoothness of $\Ln (a(x))$ it can be continued to the both
coasts $\hat{l}_\pm$ of the cut (by the well-known Plemel--Privalov's 
lemma, see
\cite{bib10}). In addition, the known equalities of Sohotsky are
fulfilled for the limiting values $K_\pm (t\pm i0)$:
\begin{align}
\label{eq3.16}
K_\pm (t\pm i0) = \frac 1{2\pi i} \bigg[ P \, \nint_0^\infty \, \frac
{\Ln (a(x)) }{x-t} \, dx \pm i\pi \Ln (a(t)) \bigg]
\end{align}
where $P \: \text{--}\!\!\!\! \int_0^\infty {\ldots} dx$ means the principal value of
regularized integral (see \cite{bib10}).

Now we can write the solution of equation \eqref{eq3.9}.

\begin{lem}\label{lem3.2}
For any complex value $\lambda = |\lambda| \exp\{i\theta\}$,
$0<\theta<2\pi$ lying outside semiaxis $0,\infty$ there exists unique
(up to a constant factor) solution of the equation \eqref{eq3.9} belonging to the class
$\mathcal{J}^*$. It has the form
\begin{align}
\label{eq3.17}
G^\lambda (w) = \frac{ w^{ (\theta - i \ln |\lambda^*|)/2\pi} }
{h(w)(w-w_-)} \exp \bigg\{ \frac 1{2\pi i} \, \nint_0^\infty \, \frac
{\Ln (a(x)) }{x-w} \, dx \bigg\} .
\end{align}
\end{lem}

 The proof of this lemma is postponed to Appendix.

As a consequence of the lemma we find for each selfadjoint extension of
the operator $\hat{M}_0$ its negative eigenvalues ($\theta = \pi$).
Namely these eigenvalues form the two-sided geometric progression
\begin{align}
\label{eq3.18}
\lambda_n^0 = \lambda_0^0 \exp \Big\{ \frac \pi{s_0} n\Big\}, \quad
n=0,\pm1,\pm2,{\ldots} 
\end{align}
where $\lambda_0^0<0$ depends on the extension.

For the proof note that any vector $\hat g \in D(\hat{M}_0^*)$ can be
represented in the form (see \cite{bib9})
\begin{align}
\label{eq3.19}
\hat g (t) = \hat f (t) + C_1 g^{\lambda=i} (t) + C_2 g^{\lambda=-i} (t)
\end{align}
where $\hat f \in D(\hat{M}_0)$ and $C_1$, $C_2$ are arbitrary
constants. In the case when
\begin{align}
\label{eq3.20}
\| g^{\lambda=i} \| = \|g^{\lambda=-i} \|
\end{align}
(and this equality is fulfilled in our case, see below)  any
selfadjoint extension of $\hat{M}_0$ is obtained by the restriction of
the operator $\hat{M}_0^*$ on the set of functions having the form
\eqref{eq3.19} with
\begin{align}
\label{eq3.21}
C_2 = \beta C_1
\end{align}
where $\beta$, $|\beta|=1$ is the parameter defining the extension (see
\cite{bib9}). We denote the extension by ${M}_0^\beta$. In particular, the
eigenfunction $g^\lambda \in D(\hat{M}_0^\beta)$ of the operator $\hat M_0^\beta$
with $\lambda<0$ has the form
\begin{align}
\label{eq3.22}
\hat{g}^\lambda (t) = \hat f (t) + C ( \hat{g}^{\lambda=i} (t) + \beta
\hat{g}^{\lambda=-i} (t) )
\end{align}
where $\hat f \in D(\hat M_0)$, $C$ depends on $\lambda$. Passing on to the
superstructures in equality \eqref{eq3.22} we get that
\begin{align}
\label{eq3.23}
\hat{G}^\lambda (w) = \hat F (w) + C ( \hat{G}^{\lambda=i} (w) + \beta 
\hat{G}^{\lambda=-i} (w) ), \quad \hat F \in \hat{\mathcal{J}} .
\end{align}
From here we obtain the following relation for the residues of the
functions $\hat{G}^\lambda$, $\hat{G}^{\lambda=\pm i}$ in poles $w_\pm$
\begin{align}
\label{eq3.24}
\res_{w_+} \hat{G}^\lambda = C (\res_{w_+}  \hat{G}^{\lambda=i} + \beta
\res_{w_+} \hat{G}^{\lambda=-i} ) ,\\
\res_{w_-} \hat{G}^{\lambda} = C (\res_{w_-} \hat{G}^{\lambda=i}
+ \beta \res_{w_-} \hat{G}^{\lambda=-i} ) . \nonumber
\end{align}
Calculating these residues with the help of \eqref{eq3.17} we get
\begin{align}
\label{eq3.25}
w_+^{1/2 - i \ln |\lambda| / 2\pi } =  C ( w_+^{1/4} + \beta w_+^{3/4}
) ,\\
w_-^{1/2 - i \ln |\lambda| / 2\pi } =  C ( w_-^{1/4} + \beta w_-^{3/4}
)  \nonumber
\end{align}
and hence
\begin{align}
\label{eq3.26}
\Big( \frac {w_+}{w_-} \Big)^{- i \ln |\lambda| / 2\pi } = \frac
{w_+^{-1/4} + \beta w_+^{1/4} }{w_-^{-1/4} + \beta w_-^{1/4} } \equiv
\gamma (\beta, s_0).
\end{align}
The ratio $w_+ / w_-$ is equal to $\exp\{ 4\pi s_0 \}$ and the value
$\gamma (\beta, s_0)$ is equal to $(1+i\beta \exp\{ \pi s_0 \} ) / (\exp\{
\pi s_0 \} + \beta i)$. It is easy to check that $|\gamma (\beta,
s_0)|=1$ i.e.\ $\gamma (\beta, s_0) = \exp \{ i \eta (\beta, s_0)\}$,
$0<\eta (\beta, s_0)< 2\pi$. Then we find that the equality
\eqref{eq3.26} is fulfilled for all $\lambda$ from the sequence
\eqref{eq3.18} and only for these $\lambda$. In \eqref{eq3.18}
$\lambda_0^0$ is equal to
\begin{align}
\label{eq3.27}
\lambda_0^0 = - \frac {e^{-\eta (\beta, s_0)}}{2s_0}.
\end{align}

\begin{lem}\label{lem3.3}
For any selfadjoint extension $\hat{M}_0^\beta$ its negative part of
spectrum consist of the eigenvalues \eqref{eq3.18}.
\end{lem}
This lemma will be proved in Appendix.

\section{Completion of the proof of Theorem \ref{thm1.1}}

Here we consider the case $m<m_1$;  assertions 2 and 3 of Theorem
\ref{thm1.1} are related to this case.

It is easily seen that any selfadjoint extension $M_0^\beta$ of $M_0$
generates the one for the operator $M_{l=1}$:
\begin{align}
\label{eq4.1}
M_{l=1}^\beta = M_0^\beta + M_1
\end{align}
and that all extensions of $M_{l=1}$ are obtained in this way.

Now we prove the existence of the sequence of the negative eigenvalues
for $M_{l=1}^\beta$ having the form \eqref{eq1.11}. Let $c=\|M_1\|$ be 
 the norm of the bounded operator $M_1$. Denote by $O_n$ the circle
in the complex plane with the center $\lambda_n^0$ and radius $2c$ and
let $\Delta_n = (\lambda_n^0 - 2c, \lambda_n^0+2c)$ be the interval
being cut on the real axis by this circle. It is obvious that for large enough
$n \geqslant n_0 > 0$ the successive intervals $\Delta_n$,
$\Delta_{n+1}$ do not intersect.

Denote
\begin{align}
\label{eq4.2}
\mathfrak{h}^0_{\Delta_n} = \int_{\Delta_n} d \mathfrak{h}^0 (\lambda)
\quad \text{and}\quad
\mathfrak{h}_{\Delta_n} = \int_{\Delta_n} d \mathfrak{h} (\lambda)
\end{align}
where $\{ \mathfrak{h}^0 (\lambda), \lambda \in \R^1\}$,
$\{ \mathfrak{h} (\lambda), \lambda \in \R^1\}$ are spectral families
of subspaces for the operators $M_0^\beta$ and $M_{l=1}^\beta$
correspondingly (see \cite{bib9}).

Let $P^0_{\Delta_n}$ and $P_{\Delta_n}$ be orthogonal
projectors on these subspaces.

From inequality which will be proved below
\begin{align}
\label{eq4.3}
\| P^0_{\Delta_n} - P_{\Delta_n} \| < 1
\end{align}
it follows that the dimensions of the subspaces
$\mathfrak{h}^0_{\Delta_n}$ and $\mathfrak{h}_{\Delta_n}$ are equal (see
\cite{bib9}). This fact and Lemma \ref{lem3.3} imply that the
operator $M_{l=1}^\beta$ has a unique simple eigenvalue $\lambda_n$ in
the interval $\Delta_n$ and $\Delta_n$ does not contain any other points of 
the spectrum of
$M_{l=1}^\beta$. In addition
\begin{align}
\label{eq4.4}
\lambda_n = \lambda_n^0 + \delta_n, \quad |\delta_n| < 2c, \quad
n\geqslant n_0
\end{align}
which gives the second assertion of  Theorem \ref{thm1.1}.

To prove the last assertion of Theorem \ref{thm1.1}, we consider the
intervals
\begin{align*}
\kappa_n = (\lambda_{n+1}^0 + 2c, \, \lambda_n^0 - 2c )
\end{align*}
lying between two successive intervals $\Delta_n$ and $\Delta_{n+1}$,
and spectral subspaces $\mathfrak{h}^0_{\kappa_n}$ and
$\mathfrak{h}_{\kappa_n}$ similar to $\mathfrak{h}^0_{\Delta_n}$, 
$\mathfrak{h}_{\Delta_n}$. From the inequality
\begin{align}
\label{eq4.5}
\| P^0_{\kappa_n} - P_{\kappa_n} \| < 1
\end{align}
it follows again that the dimensions of $\mathfrak{h}^0_{\kappa_n}$ and
$\mathfrak{h}_{\kappa_n}$ are equal. Here $P^0_{\kappa_n}$
and~$P_{\kappa_n}$ are the projectors on $\mathfrak{h}^0_{\kappa_n}$  and
$\mathfrak{h}_{\kappa_n}$ respectively. Thus due to Lemma
\ref{lem3.3} the operator $M_{l=1}^\beta$ does not have the points of
spectrum on the interval $\kappa_n$ for $n\geqslant n_0$.

\medskip
\noindent{\bf The proof of inequality \eqref{eq4.3}}

The difference $P^0_{\Delta_n} - P_{\Delta_n}$ can be represented through
the resolvents of the operators $M_0^\beta$ and $M_{l=1}^\beta$ (see
\cite{bib13})
\begin{align}
\label{eq4.6}
P^0_{\Delta_n} - P_{\Delta_n} = \frac 1{2\pi i} \int_{O_n} \big(
R_{M_0^\beta} (z) - R_{M_{l=1}^\beta} (z) \big) \, dz .
\end{align}
Use the identity
\begin{align}
\label{eq4.7}
R_{M_{l=1}^\beta} (z) = \big( E + R_{M_0^\beta} (z) M_1 \big)^{-1}
R_{M_0^\beta} (z).
\end{align}
For $z \in O_n$ the nearest point of the spectrum of $M_0^\beta$ is
$\lambda_n$ and we get
\begin{align*}
\big\| R_{M_0^\beta} (z) \big\| \leqslant \frac 1{2c}, \quad z \in O_n
\end{align*}
and hence
\begin{align*}
\big\| R_{M_0^\beta} (z) M_1 \big\| \leqslant \frac 12.
\end{align*}
Then
\begin{align*}
\big( E + R_{M_0^\beta} (z) M_1 \big)^{-1} = E + T
\end{align*}
where $T = \sum_{k=1}^\infty \big( (-1)^k \big( R_{M_0} (z) M_1 \big)
\big)^k$ and $\| T \| \leqslant 1$.

Finally
\begin{align}
\label{eq4.8}
\big\| R_{M_{l=1}^\beta} (z) - R_{M_0^\beta} (z) \big\| = \big\| T
R_{M_0^\beta} (z) \big\| \leqslant \frac 1{2c}, \quad z \in O_n .
\end{align}
Now \eqref{eq4.3} follows From  \eqref{eq4.8} and \eqref{eq4.6}.

\medskip
\noindent{\bf The proof of \eqref{eq4.5}}

For every interval $\kappa_n$, $n\geqslant n_0$ we construct the
enveloping rectangle $u_n$ with vertical sides $c$ (see Figure
\ref{fig2}).

\begin{figure}[ht]
\centering{
\begin{picture}(180,60)
\put(0,20){\line(1,0){180}}
\put(20,20){\circle{40}}
\put(20,20){\circle*{2}}
\put(20,20){\line(1,2){9}}
\put(160,20){\circle{40}}
\put(160,20){\circle*{2}}
\put(2,19){$\underbrace{\hphantom{tttttttttt}}$}
\put(142,19){$\underbrace{\hphantom{tttttttttt}}$}
\put(47,19){$\underbrace{\hphantom{ttttttttttttttttttttttt}}$}
\put(41,3){\line(0,1){30}}
\put(138,3){\line(0,1){30}}
\put(41,3){\line(1,0){97}}
\put(41,33){\line(1,0){97}}
\put(71,33){\circle*{2}}
\put(43,35){$\overbrace{\hphantom{ttttttt}}$}
\put(13,5){$\Delta_{n+1}$}
\put(11,24){$ \lambda_{n+1}$}
\put(23,44){$O_{n+1}$}
\put(53,46){$y$}
\put(80,24){$u_n$}
\put(86,6){$\kappa_n$}
\put(153,5){$\Delta_{n}$}
\put(157,24){$ \lambda_{n}$}
\put(167,42){$O_{n}$}
\end{picture}
}
\caption{\label{fig2}}
\end{figure}

As before
\begin{align}
\label{eq4.9}
\big\| P^0_{\kappa_n} - P_{\kappa_n} \big\| < \frac 1{2\pi}
\int_{\partial u_n} \big\| \big(  R_{M_0^\beta} (z) -
R_{M_{l=1}^\beta} (z) \big) \big\| \, dz .
\end{align}
For $z$ which lies on the vertical side of $u_n$, as above,
\begin{align}
\label{eq4.10}
\big\| R_{M_0^\beta} (z) - R_{M_{l=1}^\beta} (z) \big\| < \frac 1{2c}.
\end{align}
For $z$ which lies on the horizontal side of $u_n$ at the distance $y$
from the border of this side the norm of $R_{M_0^\beta} (z)$ is equal to 
$1/(2c+y)$ and
\begin{align}
\label{eq4.11}
\big\| R_{M_0^\beta} (z) - R_{M_{l=1}^\beta} (z) \big\| <
\frac{c}{(c+y)(2c+y)}
\end{align}
as it follows from the calculations similar to the previous one. From
\eqref{eq4.10} and \eqref{eq4.11} we obtain that the right part of
\eqref{eq4.9} does not exceed 
\begin{align*}
\frac 1{2\pi} \bigg( 1 + 4 \int_0^{x_n/2} \frac{c \, dy}{(c+y)(2c+y)} 
\bigg) <
\frac 5{2\pi} < 1.
\end{align*}
The inequality \eqref{eq4.5} is proved as well as Theorem \ref{thm1.1}.

\appendix

\section*{Appendix}
\renewcommand{\theequation}{A.\arabic{equation}}
\refstepcounter{section}

\subsection{The proof of Lemma \ref{lem3.1}}

Consider first the function $\hat{N}_-^* (x)$. This function has the
asymptotics for some $\alpha > 0$
\begin{align}
\label{eq5.1}
\hat{N}_-^* (x) = \begin{cases}
1+O(x^\alpha), & x\to 0, \\
1+O(x^{-\alpha}), & x\to \infty ,
\end{cases}
\end{align}
and the increment of its argument on $(0,\infty)$ is equal  to 
\begin{align}
\label{eq5.2}
\Delta_{x=0}^{x=\infty} \arg \hat{N}_-^* (x) = - 2\pi
\end{align}
Assertion  \eqref{eq5.1} follows from 
\eqref{eq2.9} and \eqref{eq2.10} for the function $N(z)$. Indeed
we see that
\begin{align}
\label{eq5.3}
N_-^* (s) = 1 + O(e^{\pm \alpha s}) \quad\text{as } s \to \mp \infty
\end{align}
for some $\alpha > 0$. After the change \eqref{eq3.4} we get \eqref{eq5.1}.

To prove \eqref{eq5.2} we note that
\begin{align}
\label{eq5.4}
\Delta_{s=-\infty}^{s=+\infty} \arg N_-^* (s) +
\Delta_{s=-\infty}^{s=+\infty} \arg N_+^* (s) = -4\pi
\end{align}
due to the principle of argument.
By \eqref{eq2.11},
\begin{align}
\label{eq5.5}
\Delta_{s=-\infty}^{s=+\infty} \arg N_-^* (s) =
\Delta_{s=-\infty}^{s=+\infty} \arg N_+^* (s) = -2\pi
\end{align}
and we obtain \eqref{eq5.2}. Thus
\begin{align}
\label{eq5.6}
\Ln \hat{N}_-^* (x) = O(x^\alpha), \quad x\to 0, \\
\Ln \hat{N}_-^* (x) = -2\pi i + O(x^{-\alpha}), \quad
x\to \infty . \nonumber
\end{align}

Now consider the function $h_+ (x) / h_- (x)$.

As before we find that
\begin{align}
\label{eq5.7}
\frac {h_+ (x)}{h_- (x)} = \begin{cases}
1 + O \Big( \frac 1{\ln x} \Big), &\quad x\to 0, \\
1 - \frac{2\pi i}{\ln x} + O \Big( \frac 1{ (\ln x)^2} \Big),
&x\to\infty,
\end{cases}
\end{align}
and
\begin{align}
\label{eq5.8}
\Delta_{x=0}^{x=\infty} \arg \frac {h_+ (x)}{h_- (x)} = 2\pi .
\end{align}
The assertion \eqref{eq5.7} follows immediately from 
\eqref{eq3.10} for $h(w)$. The equality \eqref{eq5.8} can be established by
the following representation
\begin{align}
\label{eq5.9}
\frac {h_+ (x)}{h_- (x)} = \frac {u-i/2}{u+i/2} 
\end{align}
where the function $u = u(x) = (\ln x)/2\pi - s_0$ is real and
monotonically changes from $-\infty$ to $+\infty$ when $x$
varies from $0$ to $\infty$. The linear-fractional function
$(u-i/2)/(u+i/2)$ transfers the real axis to the unit circle with 
counter-clockwise direction. This implies \eqref{eq5.8}. Thus
\begin{align}
\label{eq5.10}
\Ln \frac {h_+ (x)}{h_- (x)} = - \frac{2\pi i}{\ln x} + O \Big( \frac
1{ (\ln x)^2} \Big) + 2\pi i .
\end{align}
From \eqref{eq5.6} and \eqref{eq5.10} both assertions of
Lemma \ref{lem3.1} follow.

\subsection{The proof of Lemma \ref{lem3.2}}

We first show that the function $\hat G (w)$ (see \eqref{eq3.17})
satisfies equation \eqref{eq3.9}. Using Sohotsky's formula (see
\cite{bib11}) we get
\begin{align}
\label{eq5.11}
G^{\lambda}_{\pm} (t \pm i0) = \frac{ t \frac{\theta - i \ln
|\lambda^*| }{2 \pi} (\lambda^*)^{\eps(\pm)} }{ h_\pm (t \pm i0) (t-w_-)
} \exp \bigg\{ \frac 1{2\pi i} P \, \nint_0^{\infty} \, \frac{\Ln a(x)}{x-t} \,
dx \pm \frac 12 \Ln a(t) \bigg\}
\end{align}
where $\eps(+) = 0$, $\eps(-) = 1$. From \eqref{eq5.11} we find that
\begin{align*}
a (t) G^{\lambda}_{-} (t - i0) = \lambda^* G^{\lambda}_{+}
(t + i0) \frac{ h_+ (t + i0)}{h_- (t - i0) }
\end{align*}
and thus obtain \eqref{eq3.9}.

It follows obviously from \eqref{eq3.17} that the meromorphic function
$G^{\lambda}(w)$ has two simple poles in $w_\pm$. We prove now 
that it satisfies condition \eqref{eq3.6}.

From \eqref{eq3.17} we find that for $w = r\exp\{i \psi\} \in \Pi$
\begin{align}
\label{eq5.12}
\big| G^{\lambda}(w) \big|^2 = \frac{ r^{\theta/\pi}
|\lambda^*|^{\psi/\pi} }{ | h(w)|^2 |w-w_-|^2 } \exp \bigg\{ \frac
1{2\pi i} P \, \nint_0^{\infty} \, \Big( \frac{\Ln a(x)}{x-w} -
\frac{\Ln \bar a (x)}{x- \bar w} \Big) \, dx \bigg\}
\end{align}
where we use that $\overline{\Ln a} = \Ln \bar a$.

\begin{prop}\label{prop5.1}
For $\big| G^{\lambda}(w) \big|^2$  the following
representation is true ($w = r \exp\{ i \psi\}$)
\begin{align}
\label{eq5.13a}
\big| G^{\lambda}(w) \big|^2 &=
\frac{ (2\pi)^2 r^{\theta/\pi}
|\lambda^*|^{\psi/\pi} }{ |w-w_-|\,|w-w_+| } \\
&\ \ {}\times
\exp \bigg\{ \frac 1{2\pi
i} \int_{-\infty}^0 \big( \Ln |\hat{N}^* (s) | \big) \Big( \frac 1{s-w}
- \frac 1{s- \bar w} \Big) \, ds \bigg\} \nonumber \\
&\ \ {}\times
\begin{cases}
1 , \quad 0<\psi<\pi,\\
\displaystyle{
\frac 1{|\hat{N}^* (w) | \, |\hat{N}^* (\bar w) | }}, \quad \pi<\psi<2\pi .
\end{cases} \nonumber
\end{align}
In the case $\psi = \pi$ and $w \bar{\in} (w_-,w_+)$ the expression for 
$\bigl |G^{\lambda}(w)\bigr |^2$ reads:
\begin{align}
\label{eq5.13b}
\bigl |G^{\lambda}(w)\bigr | ^2 = \frac{ (2\pi)^2 r^{\theta/\pi} |\lambda^*| }{
|w-w_+| |w-w_-|  \hat{N}^* (w)  }.
\end{align}
\end{prop}

\medskip\noindent{\bf The proof.}

Write the expression standing in the exponent in \eqref{eq5.12} in the
form:
\begin{align}
\label{eq5.14}
\frac 1{2\pi i} &\, \bigg[ \int_0^n \Big(
\frac{ \Ln \hat{N}_-^* (x) - \Ln h_- (x) }
{x-w} - \frac{ \Ln h_- (x)}{x - \bar w} \Big) \, dx \\
& {}+
\int_0^n \Big( \frac{ \Ln \hat{N}_+^* (x) - \Ln h_+ (x) }
{x- \bar w} - \frac{ \Ln h_+ (x)}{x - w} \Big) \, dx \bigg] \nonumber \\
& {}+ 2 \Ln
(\ln n) + o(1) \quad \text{for } n\to \infty . \nonumber
\end{align}
Further we introduce a new complex variable $\zeta$ and consider in
the plane of this variable two contours $\Gamma_\pm^n$ depicted in
 Figure \ref{fig3}.

\begin{figure}[htb]
\centering{
\begin{picture}(120,120)
\qbezier(0,50)(5,0)(50,0)
\qbezier(50,0)(95,0)(100,50)
\qbezier(0,60)(5,105)(50,110)
\qbezier(50,110)(95,105)(100,60)
\qbezier(-10,50)(35,50)(50,55)
\qbezier(-10,60)(35,60)(50,55)
\qbezier(110,50)(65,50)(50,55)
\qbezier(110,60)(65,60)(50,55)
\put(48,52){$\bullet$}
\put(-20,58){\vector(4,-1){15}}
\put(120,58){\vector(-4,-1){15}}
\put(-20,60){$\bar{\cal V}$}
\put(115,60){${\cal V}$}
\put(10,60){\vector(-1,0){5}}
\put(95,60){\vector(-1,0){5}}
\put(27,51){\vector(1,0){5}}
\put(70,51){\vector(1,0){5}}
\put(86,93){\vector(1,-1){5}}
\put(87,14){\vector(-1,-1){5}}
\put(40,25){$\bullet$}
\put(40,17){$\bar w$}
\put(35,80){$\bullet$}
\put(40,80){$w$}
\put(10,40){\vector(1,4){4}}
\put(43,41){\vector(-1,3){5}}
\put(8,34){$w_+$}
\put(44,39){$w_-$}
\put(22,64){$m^n_+$}
\put(22,42){$m^n_-$}
\put(74,65){$l^n_+$}
\put(74,40){$l^n_-$}
\put(65,93){$C^n_+$}
\put(60,8){$C^n_-$}
\put(95,83){$\Gamma^n_+$}
\put(95,20){$\Gamma^n_-$}
\end{picture}
}
\caption{\label{fig3}}
\end{figure}

Each of the contours $\Gamma^n_+$ and $\Gamma^n_-$ consists of three parts:
$l^n_\pm$ are the intervals of length $n$ on the coasts of the cut $\cal V$;
$m^n_\pm$ are the same intervals on the coasts of the cut $\bar{\cal V}$;
and $C^n_\pm$ are two semicircles of radius $n$  with center at the origin.
 The contours are bypassed clockwise.

Denote by $V(\zeta,w)$
\begin{align}
\label{eq5.15}
V(\zeta,w) = \frac{\Ln \hat{N}^* (\zeta) - \Ln h(\zeta)} {\zeta-w} -
\frac{\Ln h(\zeta)}{\zeta- \bar w}
\end{align}
and rewrite the integrals in \eqref{eq5.14} in the form
\begin{align}
\label{eq5.16}
\frac 1{2\pi i} \Big[ \int_{l_-^n} V(\zeta,w) \, d\zeta +
\int_{l_+^n} V(\zeta, \bar w) \, d\zeta \Big]
\end{align}
where the interval $l_-^n$ is passed from $0$ to $n$ whence $l_+^n$ is
passed in the opposite direction. Then we decompose the sum
\eqref{eq5.16} as follows:
\begin{align}
\label{eq5.17}
\frac 1{2\pi i} &\, \Big[ \int_{\Gamma_-^n} V(\zeta,w) \, d\zeta +
\int_{\Gamma_+^n} V(\zeta, \bar w) \, d\zeta \Big] \\
=&\,
\frac 1{2\pi i} \Big[ \int_{l_-^n} \cdots d\zeta + \int_{l_+^n} \cdots
d\zeta + \int_{C_-^n} \cdots d\zeta + \int_{C_+^n} \cdots d\zeta +
\int_{m_-^n} \cdots d\zeta + \int_{m_+^n} \cdots d\zeta \Big] .
\nonumber
\end{align}
In addition the left part of \eqref{eq5.17} is equal to
the sum of residues of the integrands at the points $w=r \exp\{i\psi\}$ and
$\bar w =r \exp\{i (2\pi - \psi)\}$:
\begin{align}
\label{eq5.18}
\frac 1{2\pi i} &\, \Big[ \int_{\Gamma_-^n} \cdots \, d\zeta +
\int_{\Gamma_+^n} \cdots \, d\zeta \Big] \\
=&\, 
\begin{cases}
\Ln h(w) + \Ln h(\bar w), & 0< \psi < \pi, \\
\Ln h(w) + \Ln h(\bar w) - \Ln \hat{N}^* (w) - \Ln \hat{N}^* (\bar w),
& \pi < \psi < 2\pi .
\end{cases} \nonumber
\end{align}
It is easy to calculate that
\begin{align*}
\frac 1{2\pi i} \Big[ \int_{C_-^n} \cdots \, d\zeta +
\int_{C_+^n} \cdots \, d\zeta \Big] = 2\Ln \ln n - 2 \ln (2\pi) + o(1),
\quad n\to\infty.
\end{align*}
Thus for $\psi \neq \pi$
\begin{align}
\label{eq5.19}
\frac 1{2\pi i} &\, \Big[ \int_0^n \frac{\Ln a(x)}{x-w}\, dx - \int_0^n
\frac{\Ln a(\bar x)}{x- \bar w}\, dx \Big] + 2\Ln \ln n + o(1) \\
\nonumber
=&\,
\frac 1{2\pi i} \Big[ \int_{l_-^n} V(\zeta,w) \, d\zeta +
\int_{l_+^n} V(\zeta, \bar w) \, d\zeta \Big] +  2\Ln \ln n \\
\nonumber
=&\,
-\frac 1{2\pi i} \Big[ \int_{m_-^n} V(\zeta,w) \, d\zeta +
\int_{m_+^n} V(\zeta, \bar w) \, d\zeta \Big] + 2 \ln 2\pi \\
\nonumber
& {}+
\begin{cases}
\Ln (h(w) h(\bar w)), &0<\psi<\pi, \\
\Ln (h(w) h(\bar w)) - \Ln \big( \hat{N}^* (w) \hat{N}^* (\bar w) \big),
&\pi<\psi<2\pi.
\end{cases}
\end{align}
Passing on to the limit $n\to\infty$ we get
\begin{align*}
\frac 1{2\pi i} &\, \Big[ \, \nint_0^\infty \, \frac{\Ln a(x)}{x-w}\, dx -
\nint_0^\infty \, \frac{\Ln \bar a(x)}{x- \bar w}\, dx \Big] \\
=&\,
-\frac 1{2\pi i} \Big[ \int_{-\infty}^0  \big( V(s-i0,w) - V(s+i0,\bar
w) \big)\, ds \Big] + 2 \ln 2\pi \\
& {}+
\begin{cases}
\ln (h(w) h(\bar w)), &0<\psi<\pi, \\
\ln (h(w) h(\bar w)) - \Ln \big( \hat{N}^* (w) \hat{N}^* (\bar w) \big),
&\pi<\psi<2\pi.
\end{cases}
\end{align*}
Finally for $\psi \neq \pi$
\begin{align}
\label{eq5.20}
-\frac 1{2\pi i} &\, \bigg[ \int_{-\infty}^0  \big( V(s-i0,w) - V(s+i0,\bar
w) \big)\, ds \bigg] \\
\nonumber
&=
-\frac 1{2\pi i} \int_{-\infty}^0 \Ln \big| \hat{N}^* (s) \big| \Big(
\frac 1{s-w} - \frac 1{s-\bar w} \Big) + \frac 12 \int_{w_+}^{w_-} \Big(
\frac 1{s-w} - \frac 1{s-\bar w} \Big)\, ds .
\end{align}
This equality is obtained with the help of representation
\begin{align*}
\Ln \hat{N}^* (s \pm i0) = \Ln \big| \hat{N}^* (s) \big| + i \arg
\hat{N}^* (s \pm i0)
\end{align*}
where
\begin{align*}
\arg \hat{N}^* (s \pm i0) = \begin{cases}
0, & w_- < s < 0,\\
\pm \pi, & w_+ < s < w_-, \\
\pm 2\pi, & -\infty < s < w_+ ,
\end{cases}
\end{align*}
and similarly for $\Ln h(s\pm i0) = \ln |h(s)| = i\arg h(s\pm 0)$
\begin{align*}
\arg h(s\pm i0) = \begin{cases}
0, & w_+ < s < 0,\\
\pm \pi, & -\infty < s < w_+ .
\end{cases}
\end{align*}
In addition the terms in $(V(s-i0,w)-V(s+i0,w))$ containing
$\Ln(h(s))$ mutually cancel and the terms with $\arg \hat{N}^*
(s \pm i0)$ for $s<w_+$ cancel with the terms containing $\arg h(s\pm
i0)$. Finally
\begin{align}
\label{eq5.21}
\frac 12 \int_{w_+}^{w_-} \Big( \frac 1{s-w} - \frac 1{s-\bar w}
\Big)\, ds = \frac 12 \ln \Big( \frac{w_- - w}{w_+ - w} \Big) +
\frac 12 \ln \Big( \frac{w_- - \bar w}{w_+ - \bar w} \Big).
\end{align}
Substituting \eqref{eq5.19}, \eqref{eq5.20} and \eqref{eq5.21} into 
 \eqref{eq5.12} and noting that $\bar h (w) = h (\bar w)$ we
obtain \eqref{eq5.13a} for $\psi \neq \pi$. For the case $\psi
= \pi$ when $w = \bar w \in (-\infty,0)$ and $w \neq w_+$ 
the formula \eqref{eq5.13b} be obtained with the help of similar arguments. 
Proposition \ref{prop5.1} is proved.

\medskip
Passing on to the limits $\psi\to 0$ and $\psi\to 2\pi$ in
\eqref{eq5.13a}, \eqref{eq5.13b}
we get that
\begin{align}
\label{eq5.22a}
\big| G^\lambda_+ (t+i0) \big|^2 = |g^\lambda (t)|^2 = \frac {(2\pi)^2
t^{\theta/\pi}}{ |t-w_+| \, |t-w_-| }
\end{align}
and
\begin{align}
\label{eq5.22b}
\big| G^\lambda_- (t-i0) \big|^2 =  \frac {(2\pi)^2
t^{\theta/\pi} |\lambda^*| }{ |t-w_+| \, |t-w_-| \, | \hat{N}_-^* (t)|^2 } .
\end{align}
In particular, from \eqref{eq5.22a} we find that
\begin{align}
\label{eq5.22c}
\| g^{\lambda=i} \|^2 = \int_0^\infty \frac {(2\pi)^2 t^{1/2} \, dt
}{ |t-w_+| \, |t-w_-|t }
\end{align}
and
\begin{align}
\label{eq5.22d}
\| g^{\lambda= -i} \|^2 = \int_0^\infty \frac {(2\pi)^2 t^{3/2} \, dt
}{ |t-w_+| \, |t-w_-|t } .
\end{align}
Changing $t\to 1/t$ and taking into account that $w_+ w_-
= 1$ we achieve the equality mentioned above:
\begin{align*}
\| g^{\lambda=i} \| = \| g^{\lambda=-i} \| .
\end{align*}
Let  $U \subset \Pi$ be a bounded area containing the points
$w_\pm$. It can proved that there exists a constant $A=A(U)$ such that
for all $w \bar{\in} U$, $\bar w \bar{\in} U$,
\begin{align*}
\bigg| \int_{-\infty}^0 \Ln |\hat{N}^* (s)| 
\Big( \frac 1{s-w} - \frac 1{s-\bar w} \Big)\, ds \bigg| < A(U) .
\end{align*}
From this estimate and \eqref{eq5.13a}, \eqref{eq5.13b}
it follows that
\begin{align*}
\sup_{0 \leqslant \psi \leqslant 2\pi} \int_0^\infty \frac {
|G^\lambda(r e^{i\psi})|^2}{r} \chi_{\Pi \setminus U}
(r e^{i\psi}) \, dr < \infty, \quad U \subset \Pi
\end{align*}
i.e.\ $G^\lambda \in \hat{\mathcal{J}}^*$ for all complex $\lambda$
with $\arg \lambda \neq 0$. We now show that $G^\lambda$ is a unique
solution of the equation \eqref{eq3.9} belonging to
$\hat{\mathcal{J}}^*$. Let us assume that there is another solution
$H^\lambda \in \hat{\mathcal{J}}^*$ of equation \eqref{eq3.9}.
Consider the ratio
\begin{align}
\label{eq5.23}
\frac{H^\lambda (w)}{G^\lambda (w)} = \Phi (w)
\end{align}
(it follows from \eqref{eq5.13a}, \eqref{eq5.13b} that 
$G^\lambda (w) \neq 0$ for all $w
\neq 0$). From equation~\eqref{eq3.9} it follows that $\Phi_+(t+i0) =
\Phi_-(t-i0)$. Thus $\Phi$ is the analytic function on the complex plane
$\C^1$ with deleted point $w=0$ and can be decomposed in Loran's
series (see \cite{bib13}) 
\begin{align}
\label{eq5.24}
\Phi (w) = \sum_{n=-\infty}^{\infty} c_n w^n
\end{align}
where at least one coefficient $c_{n_0} \neq 0$ for ${n_0} \neq 0$.

Consider the ring
\begin{align*}
\mathfrak{T} = \{ 0 < \tau_0 < |w| < \tau_1 < \infty\}
\end{align*}
containing the points $w_\pm$. For large enough $R$
\begin{align*}
\min_{\tau_1 < |w| < R} \big| G^\lambda (w) \big|^2 > \frac
{\text{const}}{R^{2-\theta/\pi}}
\end{align*}
and
\begin{align*}
\min_{1/R < |w| < \tau_0} \big| G^\lambda (w) \big|^2 > \frac
{\text{const}}{R^{\theta/\pi}} .
\end{align*}
Let $U$ be an area so that $U \subset \mathfrak{T}$ and $w_\pm \in U$. Then
\begin{align}
\label{eq5.25}
\sup_{0<\psi<2\pi} &\, \int_0^\infty \frac{ \chi_{\Pi \setminus U}
(r e^{i\psi}) | H^\lambda (r e^{i\psi}) |^2 }{r} \, dr \\
\nonumber
>&\,
\frac 1{2\pi} \int_0^{2\pi} d\psi \int_0^\infty \frac{
\chi_{\Pi \setminus U} (r e^{i\psi}) | H^\lambda (r e^{i\psi}) |^2
}{r} \, dr \\
\nonumber
>&\,
\frac 1{2\pi} \min_{\tau_1 < |w| < R} \big| G^\lambda (w) \big|^2
\int_0^{2\pi} d\psi \int_{\tau_1}^R \frac{ |\Phi (r e^{i\psi})|^2 }{r}
\, dr \\
\nonumber
&{}+
\frac 1{2\pi} \min_{1/R < |w| < \tau_0} \big| G^\lambda (w) \big|^2
\int_0^{2\pi} d\psi \int_{1/R}^{\tau_0} \frac{ |\Phi (r e^{i\psi})|^2
}{r} \, dr \\
\nonumber
>&\,
\frac{\text{const}}{R^{2-\theta/\pi}} \sum_{n>0} |c_n|^2
\int_{\tau_1}^R r^{2n-1}\, dr + \frac{\text{const}}{R^{\theta/\pi}}
\sum_{n<0} |c_n|^2 \int_{1/R}^{\tau_0} r^{-2n-1}\, dr .
\end{align}
Consider two cases:

\begin{itemize}
\item[1)]
$c_{n_0} \neq 0$ for $n_0 >0$.

Then the last part of \eqref{eq5.25} can be bounded 
from below by the value $\sim R^{\theta/\pi}$.
\item[2)]
If $c_{n_0} \neq 0$ for $n_0 < 0$, \eqref{eq5.25} is bounded below by
$\sim R^{2-\theta/\pi}$.
\end{itemize}

Since $0<\theta<2\pi$, both estimates increase to infinity as 
$R\to\infty$. Consequently,
\begin{align}
\label{eq5.26}
\sup_{0<\psi<2\pi} \int_0^\infty \chi_{\Pi \setminus U}
(r e^{i\psi}) \frac{ | H^\lambda (r e^{i\psi}) |^2 }{r} \, dr = \infty .
\end{align}
Thus $H^\lambda \bar{\in} \mathcal{J}^*$ and the uniqueness of solution
$G^\lambda$ is proved. The proof of Lemma~\ref{lem3.2} is completed.

\subsection{The proof of Lemma \ref{lem3.3}}

Here we construct the resolvent of the operator $\hat{M}_0^\beta$:
\begin{align*}
R_{\hat{M}_0^\beta} (\lambda) = \big( {M}_0^\beta - \lambda E\big)^2
\end{align*}
for negative values of $\lambda = -|\lambda|$. The equation
\begin{align*}
\hat{M}_0^\beta q(t) - \lambda q(t) = f(t), \quad q\in
D(\hat{M}_0^\beta), \quad f \in L_2 \Big(\R^1, \frac{dt}{t}\Big)
\end{align*}
determining the resolvent: $q = R_{\hat{M}_0^\beta} f$ after passing to 
the superstructure $Q=Q(q)$ takes the form
\begin{align}
\label{eq5.27}
\hat{N}_-^* (t) Q_- (t-i0) - \lambda^* Q (t+i0) = f(t).
\end{align}
Introduce the function
\begin{align}
\label{eq5.28}
B^\lambda (w) &= (w-w_-) G^\lambda (w) \\
\nonumber
&=
\frac{ w^{1/2 - i \ln |\lambda^*|/2\pi}}{h(w)}
\exp \bigg\{ \frac 1{2\pi i} \, \nint_0^\infty \, \frac{ \Ln a(x)
}{x-w} \, dx \bigg\}, \quad \lambda = |\lambda| e^{i\pi}
\end{align}
which (like $G^\lambda (w)$) satisfies equation \eqref{eq3.9}.
Then we represent the solution of the
equation \eqref{eq5.27} in the form
\begin{align}
\label{eq5.29}
Q^\lambda (w) = B^\lambda (w) S^\lambda (w)
\end{align}
where the function $S^\lambda (w)$ satisfies the following equation:
\begin{align*}
S^\lambda_- (t-i0) - S^\lambda_+ (t+i0) = \frac 1{\lambda^*} \big(
B^\lambda_+ (t+i0) \big)^{-1} f(t).
\end{align*}
We choose the following solution of this equation
\begin{align}
\label{eq5.30}
S^\lambda (w) = -\frac 1{2\pi i \lambda^*} \int_0^\infty \frac {
(B^\lambda_+ (x) )^{-1} f(x) }{ x-w} \, dx + \frac{C_1}{w-w_-}
\end{align}
where $C_1$ will be determined below.

Thus the solution of equation \eqref{eq5.27} has the form
\begin{align}
\label{eq5.31a}
Q^\lambda (w) = -L^\lambda (w) + C_1 G^\lambda (w)
\end{align}
where 
\begin{align}
\label{eq5.31b}
L^\lambda (w) = \frac{ B^\lambda (w) }{2\pi i \lambda^*} \int_0^\infty
\frac {(B^\lambda_+ (x) )^{-1} f(x) }{ x-w} \, dx .
\end{align}
Using the previous calculations we find that
\begin{align}
\label{eq5.32}
\big| B^\lambda (w) \big|^2 =&\, (2\pi)^2 r |\lambda^*|^{\psi/\pi} \\
\nonumber
&{} \times
\exp
\bigg\{ \frac 1{2\pi i} \int_{-\infty}^0 \Ln |\hat{N}^* (s) | \Big(
\frac 1{s-w} - \frac 1{s-\bar w} \Big) \, ds \bigg\}
\frac{|w-w_-|}{|w-w_+|}
\end{align}
where $w=r\exp\{i\psi\}$ and $0<\psi<\pi$. A similar representation of
$|B^\lambda (w)|^2$ takes place for $\pi \leqslant \psi \leqslant
2\pi$. From \eqref{eq5.32} we find that for any bounded area $U\subset
\Pi$ containing $w_\pm$ the following inequality is true
\begin{align*}
\sup_{w \in \bar{\Pi} \setminus U} \frac{|B^\lambda (w)|^2}{r} < \infty .
\end{align*}
For $(B^\lambda (w))^{-1}$ a representation similar to \eqref{eq5.32}
takes place and  it implies
\begin{align}
\label{eq5.33}
\big(B_+^\lambda (w) \big)^{-1} f(t) \equiv b(t) \in L_2(\R^1,dt) .
\end{align}
We now show that $Q^\lambda \in \mathcal{J}^*$. Indeed from
\eqref{eq5.31a} and \eqref{eq5.31b} it is seen that $Q^\lambda$ is a 
meromorphic function with two simple poles at points $w_\pm$ and we
have to show that the condition \eqref{eq3.6} is fulfilled. It is
sufficient to check it for $L^\lambda (w)$.
\begin{align}
\label{eq5.34a}
\int_0^\infty &\, dr\, \frac{ \chi_{\bar{\Pi} \setminus U} (r e^{i\psi})
| L^\lambda (r e^{i\psi}) |^2 }{r} \\
\nonumber
&=
\int_0^\infty \chi_{\bar{\Pi} \setminus U} (r e^{i\psi}) \frac
{ | B^\lambda (r e^{i\psi}) |^2 }{r} \, \bigg| \int_0^\infty
\frac{b(x) \, dx}{x - r e^{i\psi} } \bigg|^2 \, dr \\
\nonumber
&<
\sup_{w \in \bar{\Pi} \setminus U}  \frac { | B^\lambda (w)
|^2 }{r} \int_0^\infty \bigg| \int_0^\infty \frac{b(x) \, dx }{
x - r e^{i\psi} } \bigg|^2 \, dr <
\text{const}\, \int_0^\infty dr \, \bigg| \int_0^\infty
\frac{b(x) \, dx }{ x - r e^{i\psi} } \bigg|^2 .
\end{align}

\begin{prop}\label{prop5.2}
For any function $b\in L_2(\R^1,dt)$
\begin{align}
\label{eq5.34b}
\sup_\psi \int_0^\infty dr \, \bigg| \int_0^\infty \frac{b(x) \, dx }{
x - r e^{i\psi} } \bigg|^2 < \infty .
\end{align}
\end{prop}

{\bf The proof.}
Write 
\begin{align}
\label{eq5.35}
\int_0^\infty dr \bigg| \int \frac{b(x) \, dx }{ x - r e^{i\psi} } \bigg| =
\int_0^\infty dr \int_0^\infty \int_0^\infty
\frac{b(x) \bar{b}(y) \, dx \, dy }{ (x - r e^{i\psi})(y- r e^{i\psi}) }\, .
\end{align}
We can easily calculate that
\begin{align}
\label{eq5.36}
\int_0^\infty \frac{dr}{(x - r e^{i\psi})(y- r e^{-i\psi}) } =
-\frac{ \ln (y/x) - 2(\pi-\psi)i }{ y e^{i\psi} - x e^{-i\psi} }
\end{align}
and thus \eqref{eq5.35} is equal to
\begin{align*}
- \int_0^\infty \int_0^\infty \frac{ \ln (y/x) - 2(\pi-\psi)i }
{ y e^{i\psi} - x e^{-i\psi} } \, b(x) \bar{b} (y) \, dx\,dy .
\end{align*}
After unitary Mellin's transformation 
\begin{align*}
\omega': L_2(\R^1_+) \to L_2(\R^1), \quad
b(x) \longrightarrow \tilde{b} (s) = \frac 1{\sqrt{2\pi}}
\int_0^\infty x^{-is-1/2} b(x)\, dx
\end{align*}
\eqref{eq5.35} reads
\begin{align}
\label{eq5.37}
-\frac 1{2\pi} \int_{-\infty}^\infty \big| \tilde{b} (s) \big|^2 \, ds
\int_0^\infty \frac{ \ln \xi - 2(\pi-\psi)i }{ \xi e^{i\psi} -
e^{-i\psi} } \, \xi^{-is-1/2} \, d\xi .
\end{align}
It is easy to calculate that
\begin{align}
\label{eq5.38}
-\frac 1{2\pi} \int_0^\infty \frac{ \ln \xi - 2(\pi-\psi)i }
{ \xi e^{i\psi} - e^{-i\psi} } \xi^{-is-1/2} \, d\xi =
\frac{2\pi}{ (e^{-2\pi s} +1) e^{2\psi s} } .
\end{align}
Since the right-hand side of \eqref{eq5.38} is bounded, 
\eqref{eq5.34b}  follows from \eqref{eq5.37} and \eqref{eq5.38}.
 The proposition is proved and hence $Q^\lambda \in
\mathcal{J}^*$.

Since the function $q^\lambda (t) = Q_+^\lambda (t+i0)$ belongs to
$D(\hat{M}_0^\beta)$, it admits  representation similar to \eqref{eq3.19}
together with \eqref{eq3.21}. Going over to the
superstructures we obtain
\begin{align}
\label{eq5.39}
Q^\lambda (w) = F (w) + C_0 \big( G^{\lambda = i} +
\beta G^{\lambda = -i} \big)
\end{align}
where $F \in \hat{\mathcal{J}}$. From this and \eqref{eq5.31a} we get
the following relations for the residues of
the functions $Q^\lambda$, $L^\lambda$, $G^{\lambda =\pm i}$ and
$G^{\lambda}$, at the points $w_\pm$:
\begin{align*}
\res_{w_\pm} Q^\lambda = C_1 \res_{w_\pm} G^{\lambda} + \res_{w_\pm}
L^\lambda = C_0 \big( \res_{w_\pm} G^{\lambda = i} + \beta
\res_{w_\pm}  G^{\lambda = -i} \big) .
\end{align*}
Thus we get the following equations w.r.t.\ constants $C_1$, $C_0$:
\begin{align}
\label{eq5.40}
C_1 \res_{w_+} G^\lambda - C_0 \big( \res_{w_+} G^{\lambda = i} +
\beta \res_{w_+} G^{\lambda = -i} \big) = -\res_{w_+} L^\lambda , \\
C_1 \res_{w_-} G^\lambda - C_0 \big( \res_{w_-} G^{\lambda = i} +
\beta \res_{w_-} G^{\lambda = -i} \big) = -\res_{w_-} L^\lambda .
\nonumber
\end{align}
We see that the resolvent $R_{\hat{M}_0^\beta} (\lambda)$ exists
iff the system \eqref{eq5.40} is solvable for any right part or in
other words for those $\lambda < 0$ for which the homogeneous system
has only trivial solution: $C_1 = C_0 = 0$. The other $\lambda$'s, 
$\lambda < 0$,
form the negative spectrum of the operator $\hat{M}_0^\beta$ and for
these $\lambda$ the homogeneous system has non-zero solution $(C_1,
C_0)$. One can check that in this case $C_1 \neq 0$ and the homogeneous
system is equivalent to the system \eqref{eq3.24}, which has a 
solution iff $\lambda$ belongs to the sequence \eqref{eq3.18}. Thus 
Lemma \ref{lem3.3} is proved.

\section*{Concluding remarks}

{\bf 1.}
We have established Efimov's effect ($\lambda_n \to 0$, $n \to
- \infty$), Thomas's  effect ($\lambda_n \to -\infty$, $n \to \infty$)
for the operator ${M}_0^\beta$. As we see it implies Thomas's  effect
for the operator ${M}_{l=1}$. However nothing is known about Efimov's 
effect for that operator.

\medskip\noindent{\bf 2.}
The methods of this work (as well as \cite{bib1}) can be used
for the investigation of the operators $T_l$, $l>1$, if we know 
the number and the position of zeros for the function $N_l(z)$, $z\in I$.

\section*{Acknowledgment}
This work is supported by RFBR, grant 11-01-00485a.


\end{document}